\documentstyle[epsf]{aipproc}

\newcommand{\beq}{\begin{equation}}
\newcommand{\eeq}{\end{equation}}
\def\lsim{\mathrel{\rlap{\lower 3pt\hbox{$\mathchar"218$}}
     \raise 2.0pt\hbox{$\mathchar"13C$}}}
\def\gsim{\mathrel{\rlap{\lower 3pt\hbox{$\mathchar"218$}}
     \raise 2.0pt\hbox{$\mathchar"13E$}}}

\begin{document}

\title{An Ultraviolet Redshift Survey: \\ A New Estimate of the Local \\ Star Formation Rate }
\author{M.A. Treyer$^1$, R.S. Ellis$^2$, B. Milliard$^3$, J. Donas$^3$}

\address{$^1$ Astrophysikalisches Institut Potsdam, Germany \\
$^2$ Institute of Astronomy, Cambridge, UK \\
$^3$ Laboratoire d'Astronomie Spatiale, Marseille, France}

\maketitle

\begin{abstract}
We present the first results of an ongoing spectroscopic survey of
galaxies selected in the rest-frame ultraviolet. The source catalogue
was constructed from a flux-limited sample of stars, galaxies and QSOs
imaged at 2000 \AA\ with the FOCA balloon-borne camera (\cite{code6}).
The redshift distribution obtained for 45 galaxies spans $0<z<0.3$ and
a high fraction of the spectra show intense nebular emission lines and
ultraviolet-optical colors considerably bluer than can be accounted for
via normal Hubble sequence galaxies. From the rest-frame ultraviolet
galaxy luminosity function, and adopting a normal IMF, we use the
integrated UV light to our survey limit to estimate the local volume
averaged star formation rate. We find a value significantly larger than
recent estimates and argue this must be a lower limit to the true
value. Our results suggest the local abundance of star-forming galaxies
has been underestimated in surveys based on optical data.
\end{abstract}

\section*{Introduction}

Considerable progress has recently been made in delineating the star
formation history of normal field galaxies (see \cite{code1} for a
recent review).  However, it remains unclear whether the various star
formation rates (SFR) are being inferred consistently.  Various
diagnostics have been used including nebular emission lines
(\cite{code2}, \cite{code3}) and the far UV continuum flux
(\cite{code4}, \cite{code5}).  In most cases, these estimates are based
on only the bright portion of the luminosity function.

The existing and forthcoming data could be better interpreted if there
was a greater overlap between the low and high redshift diagnostics.
With this in mind, we are undertaking a deep UV-selected galaxy
redshift survey based on FOCA balloon-borne images (\cite{code6}). Our
survey is designed to span a moderate redshift range 0$<z<$0.3 which
aims to provide an independent measure of the declining SFR and a robust
estimate of the present star formation density.

\section*{Data}

Our survey strategy can be summarized as follows:
\begin{itemize}
\item The most suitable field for study from Milliard et al's series 
of 4 UV exposures is Selected Area 57 (SA57): it has been studied
with both the FOCA 1000 and FOCA 1500 instruments, ensuring the 
deepest, most reliable wide-field (2.3$^{\circ}$) catalogue. 
\item The limiting magnitude for reliable photometry at 2000\AA\ 
is $m_{UV}$=18.5 at which the surface density of galaxies is 
$\simeq$ 200 deg$^{-2}$.
\item The imaging resolution of FOCA data is 20 arcsec which provides 
a positional accuracy of $\simeq$ 4.5 arcsec rms. Target astrometry 
suitable for multi-object spectroscopy has been obtained by matching 
the FOCA 1000 catalogue for SA57 with APM scans of the Palomar Sky 
Survey 103a-O and 103a-E plates. 
\item Optical spectroscopy has been conducted with the Hydra multi-fiber 
spectrograph on the 3.5m WYIN telescope within a 1$^{\circ}$ field
and work is continuing with the WYFFOS fiber spectrograph at the prime 
focus of the 4.2m WHT. 
\item Reliable spectra have so far been analysed for 45 sources 
to $m_{UV}$=18.5, of which 3 are QSOs and 2 are stars.
\end{itemize}

%
%
\begin{figure}[t]
\unitlength1cm
\begin{picture}(14.5,14.5)  
{\epsfxsize=14.5cm 
\epsfbox{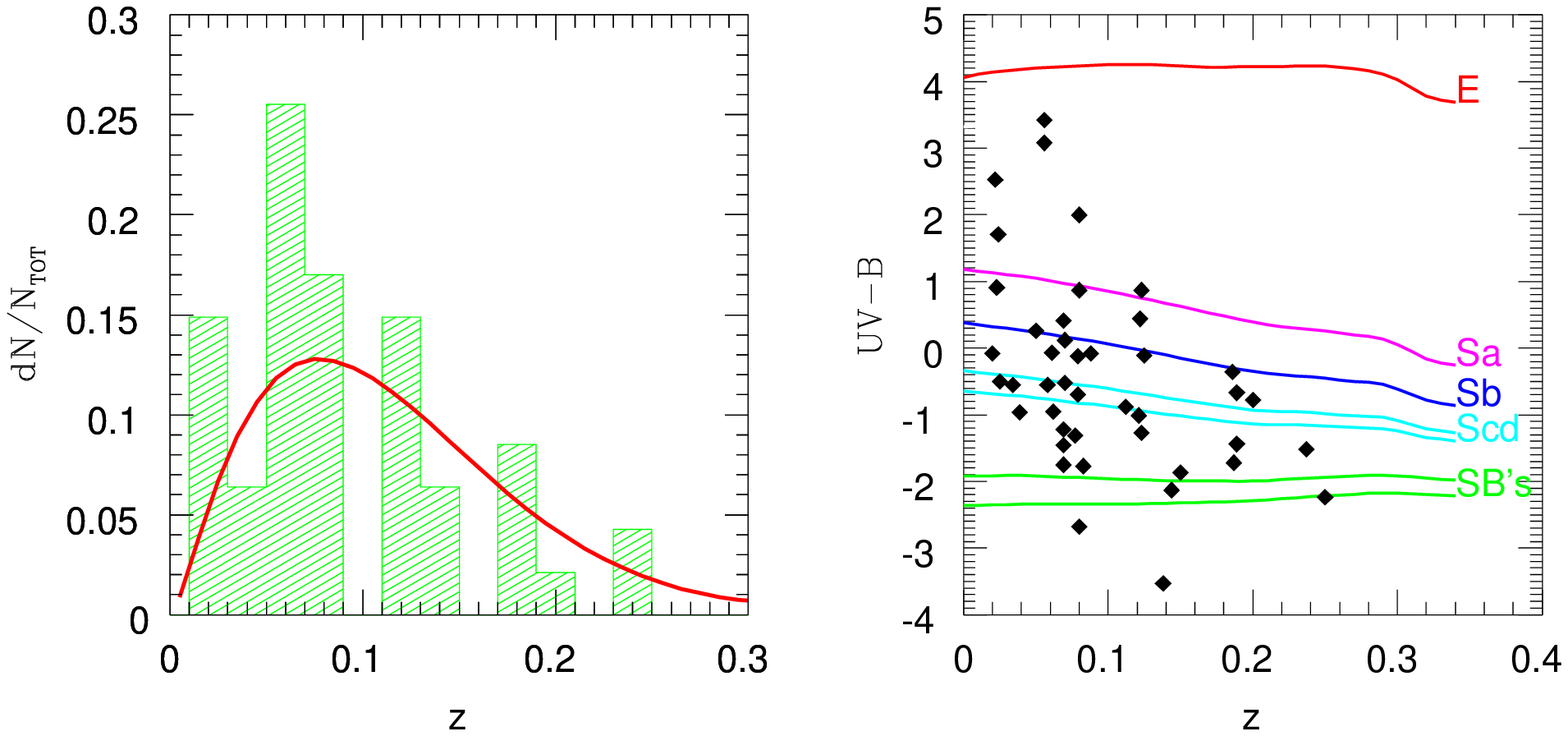}}   
\end{picture} 
\vskip -6.cm
\caption{Normalized redshift distribution for the WIYN results.
The solid line is the predicted distribution assuming the King \& Ellis
(1985) type-dependent optical luminosity function and $k$-corrections 
derived from model spectra of Poggianti (1997).}
\caption{Ultraviolet-B color versus redshift. Curves refer to the predicted
color-redshift relation for the Poggianti (1997) model spectra and 
are used to assign $k$-corrections to each source. SB's are two 
starburst models described in the text.}
\end{figure}
%

%
\begin{figure}[t]
\unitlength1cm
\begin{picture}(14.5,14.5)  
{\epsfxsize=14.5cm
\epsfbox{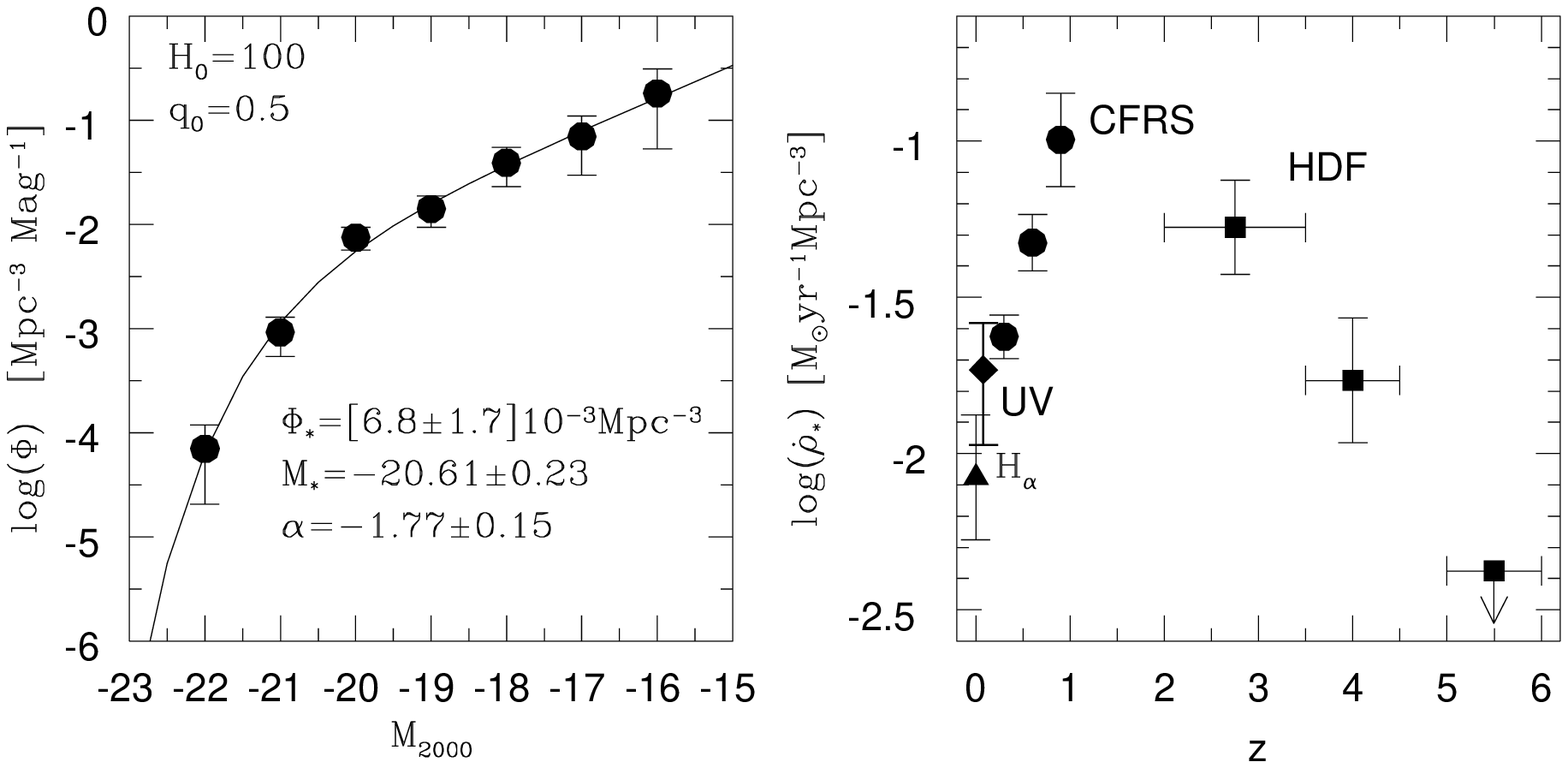}}   
\end{picture}
\vskip -6.cm
\caption{The observed UV luminosity function ($H_o$=100 kms sec$^{-1}$
Mpc$^{-1}$) and a Schechter function fit whose parameters are indicated.}
\caption{The redshift dependence of the comoving volume-averaged star 
formation rate following Madau (1997) and assuming a Salpeter IMF.
Our estimate is the diamond symbol placed at a mean redshift 
$\bar z \sim 0.15$. The $H_{\alpha}$ data is from Gallego et al. 
(1995), the CFRS data from Lilly et al. (1996) and the HDF data from 
Madau (1997).}
\end{figure}

%
\section*{Results}
Figure 1 shows the redshift distribution obtained from our current
sample. A high fraction of the spectra reveal strong [O II] emission
lines. We have predicted UV-optical colors by folding model spectra
from Poggianti (1997) \cite{code8} through the filter functions of both
the UV ($\Delta\lambda$=150 \AA\ ) and photographic photometric
systems. The UV-optical colors (Figure 2) are considerably bluer than
can be accounted for via normal Hubble sequence galaxies (\cite{code7})
as would be expected if the bulk of the UV-selected sources are
star-forming galaxies. The two starburst (SB) models in Figure 2 refer
to models constructed assuming a $10^7$ year long burst of
star-formation prior to the present time involving 30 \% and 80 \% of
the galactic mass respectively.  Extinction is $not$ included in these
models.

By assigning $k$-corrections on the basis of the color-redshift
relation, we have also derived a UV luminosity function shown in Figure 3.
Our best fit Schechter function has parameters indicated on the figure
and has a surprisingly steep faint end slope. 
The luminosity density at 
2000\AA\ derived from the integrated emission of the {\it observed} 
galaxies, with absolute magnitudes $M_{UV}\lsim -16$ and mean redshift 
$\bar z \sim 0.15$, is:

\beq
\rho_{2000}= (1.6\pm 0.7) \times 10^{26}  ~h~{\rm ergs~s^{-1}
Hz^{-1} Mpc^{-3}}.
\eeq
\noindent The ultraviolet radiation flux can be used to indicate 
the instantaneous ejection rate of heavy element $\dot\rho_Z$
(\cite{code9}):
\beq
{\rho_{2000} \over h~ {\rm ergs~s^{-1} Hz^{-1} Mpc^{-3}}} 
\approx 3.8\times 10^{29}
{\dot\rho_Z \over h~ {\rm M_{\odot} yr^{-1}Mpc^{-3}}}
\eeq
\noindent The conversion efficiency is fairly insensitive to the
assumed initial mass function (IMF), unlike that for the star-formation-rate $\dot\rho_\star$.
Assuming a Salpeter IMF:

\beq
\dot\rho_\star = 42\times \dot\rho_Z \approx (1.8\pm 0.7) \times 10^{-2}h~ 
{\rm M_{\odot} yr^{-1}Mpc^{-3}}.
\eeq
This value is shown in Figure 4 together with other recently published
estimates as a function of redshift (\cite{code9}). Our estimate
appears to be twice as large as that derived from $H_\alpha$ surveys
(\cite{code2}). As we have not taken dust extinction into account and
our luminosity function is rising steeply at the faint end, the true 
integrated value could be even larger.

In summary, from various viewpoints, our data suggests that optical
surveys of the local universe may have seriously underestimated the
abundance of star-forming galaxies. If SA57 is a representative field, 
our result reduces quite significantly the claimed redshift evolution of 
the star-formation-rate inferred from the steep slope of the faint blue 
galaxy counts, a substantial fraction of which are being sampled 
at 2000 \AA\ at modest redshift.


\end{document}